# Experimental Observation of Dirac Nodal Links in Centrosymmetric Semimetal TiB$_2$

Zhonghao Liu,[1,†] Rui Lou,[2,†] Pengjie Guo,[2] Qi Wang,[2] Shanshan Sun,[2] Chenghe Li,[2] Setti Thirupathaiah,[3,4] Alexander Fedorov,[3,5] Dawei Shen,[1] Kai Liu,[2] Hechang Lei,[2,*] Shancai Wang[2,*]

[1] *State Key Laboratory of Functional Materials for Informatics and Center for Excellence in Superconducting Electronics, SIMIT, Chinese Academy of Sciences, Shanghai 200050, China*

[2] *Department of Physics and Beijing Key Laboratory of Opto-electronic Functional Materials & Micro-nano Devices, Renmin University of China, Beijing 100872, China*

[3] *Institute for Solid State Research, IFW Dresden, D-01171 Dresden, Germany*

[4] *Solid State and Structural Chemistry Unit, Indian Institute of Science, Bangalore, Karnataka 560012, India*

[5] *Department of Physics, St. Petersburg State University, St. Petersburg 198504, Russia*

[†] These two authors contributed equally to this work.
[*] Corresponding authors: hlei@ruc.edu.cn & scw@ruc.edu.cn

The topological nodal-line semimetal state serving as a fertile ground for various topological quantum phases, where topological insulator, Dirac semimetal, or Weyl semimetal can be realized when the certain protecting symmetry is broken, has only been experimentally studied in very few materials. In contrast to discrete nodes, nodal lines with rich topological configurations can lead to more unusual transport phenomena. Utilizing angle-resolved photoemission spectroscopy and first-principles calculations, here we provide compelling evidence of nodal-line fermions in centrosymmetric semimetal TiB$_2$ with negligible spin-orbit coupling effect. With the band crossings just below the Fermi energy, two groups of Dirac **nodal rings are clearly observed without any interference from other bands, one surrounding the Brillouin zone (BZ) corner in the horizontal mirror plane $\sigma_h$ and the other surrounding the BZ center in the vertical mirror plane $\sigma_v$. The linear dispersions forming Dirac nodal rings is as wide as 2 eV. We further observe that the two groups of nodal rings link together along the $\Gamma$-$K$ direction, composing a nodal-link configuration. The simple electronic structure with Dirac nodal links mainly constituting the Fermi surfaces suggests TiB$_2$ a remarkable platform for studying and applying the novel physical properties related to nodal-line fermions.**

## Introduction

Topological materials with symmetry protected nodes have recently attracted great attentions in condensed matter physics. Perturbation that preserves a certain symmetry cannot remove the nodes by opening a full direct gap in these materials. When such nodes are close to the Fermi level ($E_F$), the low-energy quasiparticle excitations are drastically different from the usual Schrödinger-type fermions. These nodes can be classified by their dimensionality [1, 2]. The zero-dimensional ones include Dirac points, Weyl points, 3-fold degenerate points and other higher-fold degenerate nodal points. With the 4- or 2-fold degenerate Dirac or Weyl points, respectively, Dirac and Weyl semimetals have been theoretically predicted and experimentally confirmed [3-12]. More recently, 3-fold degenerate fermions, which conceptually lie between Dirac fermions and Weyl fermions, have been demonstrated to exist in the ZrTe family of compounds [13, 14].

Nodal ring, nodal link, and nodal chain belong to the one-dimensional (1D) nodal-line systems whose nodes extend along 1D lines instead of discrete points in the three-dimensional (3D) Brillouin zone (BZ) [2, 15-19]. Protected by the combination of spatial-inversion symmetry and time-reversal symmetry (the $P·T$ symmetry) or certain crystalline symmetry [20, 21], the nodal line may either take the form of an extended line running across the BZ, whose ends meet at the BZ boundary, or wind into a closed loop inside the BZ, or even form a chain consisting of several connected loops.

Nontrivial Berry phase around the nodal lines would shift the Landau-level index by 1/2 [17] and lead to drumhead surface states (SSs) [18, 19].

Although band theory has predicted the existence of nodal-line fermions in some materials [19], only very few candidates have been experimentally studied by angle-resolved photoemission spectroscopy (ARPES), *i.e.*, CaAgAs, PbTaSe$_2$, and the ZrSiS family [20, 22-26]. Very recently, two AlB$_2$-type diborides TiB$_2$ and ZrB$_2$ have been proposed to possess nodal-line configurations coexisting with a pair of the triply degenerate nodal points (TNPs) with negligible spin-orbit coupling (SOC) [27, 28]. The formation of the TNPs was well uncovered by calculations, but these nodes cannot be observed by ARPES due to locating above $E_F$. On the other hand, the nodal lines could be clearly observed with the crossing located below $E_F$. In this work, we unambiguously identify two groups of Dirac nodal rings in TiB$_2$ by means of ARPES and first-principles calculations. The nodal rings embedded in the mirror plane $\sigma_h$ (the $\Gamma$-$K$-$M$ plane) around $K$ points compose one group and those in the mirror plane $\sigma_v$ (the $\Gamma$-$A$-$H$ plane) around $\Gamma$ point compose the other. Without interfering with other bands, the dispersions forming the nodal rings exhibit linearly in a wide energy range of ~2 eV. These two groups of nodal rings linked together along the $\Gamma$-$K$ direction forming a nodal-link configuration, which goes beyond the isolated nodal line configuration in other systems. The compelling evidences of the nodal-line fermions existing just below $E_F$ and the nodal links mainly constituting the Fermi surfaces (FSs) provide an ideal system for further investigations and potential applications on transport phenomena of nodal lines.

**Results and Discussion**

TiB$_2$ has a simple AlB$_2$-type centrosymmetric structure with the space group *P6/mmm* (No. 191) [29]. As shown in Fig. 1a, titanium and boron atoms lie in planar close-packed hexagonal layers alternately. Figure 1b shows the x-ray diffraction (XRD) pattern of a TiB$_2$ crystal, indicating the measured surface is the (00*l*) plane. The mirror-like (001) surface is illustrated in the inset of Fig. 1b. In Fig. 1c, the zero-field resistivity $\rho_{xx}$ of TiB$_2$ exhibits a metallic behavior in the measured temperature range from 3 K to 300 K, and the magnetic field dependence of Hall resistivity $\rho_{xy}$ at $T$ = 3 K is displayed in the inset of Fig. 1c. Based on the nonlinear feature of $\rho_{xy}$ extending to high fields, electrons and holes are believed to coexist near $E_F$ in TiB$_2$. Thus, we perform a quantitative analysis by fitting $\rho_{xy}$ with the two-carrier model (superimposed on the original data) [30]. The extracted carrier densities of TiB$_2$ are $n_e$ = 2.74(4) × 10$^{21}$ cm$^{-3}$ and $n_h$ = 2.63(4) × 10$^{21}$ cm$^{-3}$, respectively. They are much higher than those of Dirac semimetals like Na$_3$Bi (~10$^{17}$ cm$^{-3}$ [31]) and Cd$_3$As$_2$ (~10$^{18}$ cm$^{-3}$ [32]) with discrete nodes, but comparable to that of nodal-line semimetals ZrSi$M$ ($M$ = S, Se, and Te) ($n_e$, $n_h$ ~ 10$^{20}$ cm$^{-3}$ [33, 34]). The concentrations in TiB$_2$ estimated from first-principles calculations ($n_e$, $n_h$ = 1.04, 1.10 × 10$^{21}$ cm$^{-3}$) also show good consistency with experimental results.

The bulk BZ and high-symmetry points are indicated in Fig. 1d. The Γ-$K$-$M$ and Γ-$A$-$H$ planes are two mirror planes ($\sigma_h$ and $\sigma_v$) of $D_{6h}$ group, respectively. Figures 1e and 1f show the ARPES intensity plot in a wide energy region along the Γ-$K$-$M$ direction and the corresponding second derivative plot, respectively. The calculated band dispersions have been superimposed on the second derivative plot, indicating the rather good agreement between the calculations and experimental data. Obviously, two linear band-crossing features denoted as $\alpha$ and $\beta$ can be identified in Fig. 1f.

To comprehensively and accurately investigate the electronic structure of TiB$_2$, we have scanned large portions of $k$-space using various photon energies. The measured and calculated FSs are shown in Figs. 2a-2d. Combining with the calculated band structure along high-symmetry lines, as displayed in Fig. 2g, one can distinguish every FS in the whole 3D FS map of Fig. 2c. Besides an electron-like and a hole-like FSs around $A$ point in the $k_z$ = π plane, the overall FSs are mainly composed of the nodal rings, with the nodal rings ($r_1$) embedded in the $\sigma_h$ plane around $K$ points linked with the ones ($r_2$) in the $\sigma_v$ planes around Γ point at $J$ points along the Γ-$K$ directions. Figure 2d shows the top view of the calculated 3D FSs, namely the two-dimensional (2D) FSs projected on the (001) surface. With the FSs centered at $A$ point, the $r_1$ nodal ring surrounding $K$ point, the projection of $r_2$ nodal ring around $\Gamma$ point, and the $J$ nodal-link point all clearly resolved in Fig. 2a, the experimental map taken with the photon energy of 80 eV well reproduces the feature of the 2D projected FSs. Due to the ARPES spectra reflects the electronic states integrated over a certain $k_z$ region of bulk BZ and the electronic states at $k_z = 0$ and $\pi$ have main contributions [24, 35], the $r_1$ nodal ring and the FSs centered at $A$ point could be prominently distinguished in a certain range of photon energies. The banana-shaped $r_1$ nodal ring becomes a hole-like pocket, when moving towards high binding energy, as shown in Fig. 2b.

In Fig. 2g, one can find four band-crossing features near $E_F$ along the $M$-$K$, $K$-$\Gamma$, $H$-$\Gamma$, and $\Gamma$-$A$ directions, namely $\alpha$, $\beta$, $\gamma$, and $\delta$, respectively. The nodes of the $\alpha$ and $\beta$ band-crossing features locate below $E_F$, thus the whole $r_1$ nodal rings are resolved in the FS map. The nodes of the $\gamma$ and $\delta$ band-crossing features are below and above $E_F$, respectively, leading to the discontinuity of $r_2$ nodal rings near $A$ point, as illustrated in Figs. 2c and 2d. The $\delta$ is a TNP, which is a crossing by a doubly degenerate band and a nondegenerate band as the presence of the $C_{6v}$ symmetry group for the $\Gamma$–$A$ line. Due to time reversal symmetry, a pair of the TNPs should locate at the two sides of the $\Gamma$ point. Further, the SOC introduces small gaps of 22, 25, 22 and 33 meV to the nodes of $\alpha$, $\beta$, $\gamma$, and $\delta$, respectively (see the calculated bulk band structure without SOC in Fig. S1 of the Supplementary Materials (SM)). The gap sizes are much less than that in CaAgAs (~73 meV) [22] and ZrSiTe (~60 meV) [26], and comparable to that in ZrSiS and ZrSiSe (~25-35 meV) [20, 26], indicating the weak SOC effect on the bulk electronic structure of TiB$_2$. The $r_2$ nodal ring in the $\sigma_v$ plane formed by $\gamma$ and $\delta$ needs to be further demonstrated by the photon-energy-dependent studies, which will be discussed later. Here, we extract the Fermi wave vectors of $r_1$ according to the ARPES data and plot them in Fig. 2e with superimposed calculation contours. The size and shape of the calculated ones are well consistent with those determined by ARPES.

To quantitatively study the feature of $r_1$ nodal ring embedded in the $\sigma_h$ plane, we record the ARPES spectra near $E_F$ along the $M$-$K$ and $\Gamma$-$K$ directions, as indicated by cuts 1 and 2 in Fig. 2f. The intensity plots and corresponding second derivative plots are presented in Figs. 3a-3d. The overall band structures agree well with the appended theoretical calculations. We also notice that some band dispersions near $\beta$ along the Γ-$K$ direction are not reproduced by bulk calculations. These bands most likely come from the contributions of bulk states in other $k_z$ planes due to the $k_z$ broadening effect [24, 35].

To avoid the interference with other bands, we use the photon energy of 110 eV, in principle corresponding to the Γ-$K$-$M$ plane, to record the intensity. As shown in Figs. 3e and 3f, the $\alpha$ and $\beta$ band-crossing features are more outstandingly recognized. The corresponding momentum distribution curves (MDCs) in Figs. 3g and 3h unambiguously demonstrate the linear dispersions in a large energy range, and the nodes of $\alpha$ and $\beta$ are below $E_F$. The sizes of $\alpha$ and $\beta$ FSs along the Γ-$K$ direction are 0.11 and 0.07 π/a', respectively. The result is in agreement with the data extracted from de Haas-van Alphen effect measurements [36]. To quantitatively determine the energy gaps at the nodes, we extract the energy distribution curve (EDC) at the center (about -0.81 π/a') of $\beta$, as illustrated in Fig. 3i. By zooming in the near-$E_F$ area, one can clearly see two peaks in the inset of Fig. 3i. We use two Lorentzian curves to fit this single EDC and superimpose the fitting result, as blue curve. The gap of ~25 meV at the node of $\beta$ can be obtained, showing good consistency with theoretical calculation. Although the faint intensity due to the matrix element effect obstructs the determination of band gap at $\alpha$, we can reasonably expect a small gap opening based on the high consistency between experiments and calculations.

As mentioned above, the $r_2$ nodal ring embedded in the $\sigma_v$ plane needs to be further demonstrated in the $k_z$-$k_{//}$ plane. We perform $k_z$-dependent measurements by varying the photon energy from 50 to 124 eV, covering more than one BZ along the $k_z$ direction. Figure 4a shows the intensity plot at $E_F$ as a function of photon energy and $k_{//}$, which is oriented along the Γ-$K$ ($A$-$H$) direction. One can see that the photoemission intensity is enhanced around $A$ point (86 eV) and suppressed around Γ point (110 eV) along the $A$-Γ direction, showing a periodic modulation along with photon energies. By comparing with the projections of calculated FSs on the Γ-$A$-$H$ plane in Fig. 4b, the $r_2$ nodal rings passing through the *J* nodal-link point are traced out by open circles in Fig. 4a. With the observation of both the (001) projection and 3D character of $r_2$ nodal ring in Figs. 2a and 4a, respectively, the existence of nodal rings embedded in the $\sigma_v$ planes can be clearly proved. Thus, the nodal-link structure formed by $r_1$ and $r_2$ nodal rings is confirmed.

Figure 4c shows the ARPES intensity plot along the *A-H* line taken by the photon energy of 86 eV. As compared with the band structure along the Γ-*K* direction in Fig. 3e, an electron-like band and a hole-like band cross $E_F$ around *A* point while no band exist at Γ point. Moreover, the *β* band-crossing feature along the Γ-*K* direction disappears along the *A-H* direction. These contrasts reveal the 3D nature of the band structure around *A* point and the embedding of $r_1$ nodal ring in the $\sigma_h$ plane. Figures 4d-4g show the constant-energy maps in the $k_x$-$k_y$ plane, which are taken by the photon energies of 110 eV ($k_z \sim 0$) and 120 eV ($k_z \sim 0.4\pi/c$), respectively. As illustrated in Fig. 4f, the nodal-link configurations of $r_1$ and $r_2$ are also recognized like that in Fig. 2a.

Besides the bulk evidence of nodal-line, the drumhead SSs appearing inside the projections of the nodal lines serve as the surface signature. As discussed above, small gaps will open at the nodes of *α, β, γ,* and *δ,* when including SOC effect. Since the presence of inversion center in TiB$_2$, we calculate the product of the parities of the occupied states, which are all doubly degenerate due to the *P·T* symmetry, at the eight time-reversal invariant momenta with SOC [37]. The results are listed in Table S1 of the SM. With a $Z_2$ index as (0; 001), the SSs of TiB$_2$ predicted in Refs. [27] and [28] are not topologically protected, a small perturbation on the surface could hinder the observation of SSs in our experiments.

## Conclusion

By systematically mapping out the 3D electronic structure both in experiment and theory, we unambiguously demonstrate the existence of nodal-line fermions in TiB$_2$, hosting Dirac nodal links formed by two groups of nodal rings with negligible SOC effect, under the protection of *P·T* symmetry and certain mirror reflection symmetries.

The fact that the nodal-line fermions existing just below $E_F$ combined with the nodal links mainly constituting the FSs must have significant contribution to the low-energy excitations, which are favorable to the emergence of associated novel transport properties. The simple electronic structure composed of large energy range of linearly dispersive bands offers a good platform for further studies into Dirac physics.

## Acknowledgements

We thank Zhong-Yi Lu for helpful discussions. The work was supported by the National Key R&D Program of China (Grants No. 2016YFA0300504 and No. 2017YFA0302903), and the National Natural Science Foundation of China (Grants No. 11774421, No. 11574394, No. 11774423, No. 11774424, No. 11227902, and No. 11704394). R.L., K.L. and H.L. were supported by the Fundamental Research Funds for the Central Universities, and the Research Funds of Renmin University of China (RUC) (Grants No. 17XNH055, No. 14XNLQ03, No. 15XNLF06, and No. 15XNLQ07). Z.L. acknowledges Shanghai Sailing Program (No. 17YF1422900). A.F. acknowledges the support of Saint Petersburg State University (Grant No. 15.61.202.2015).

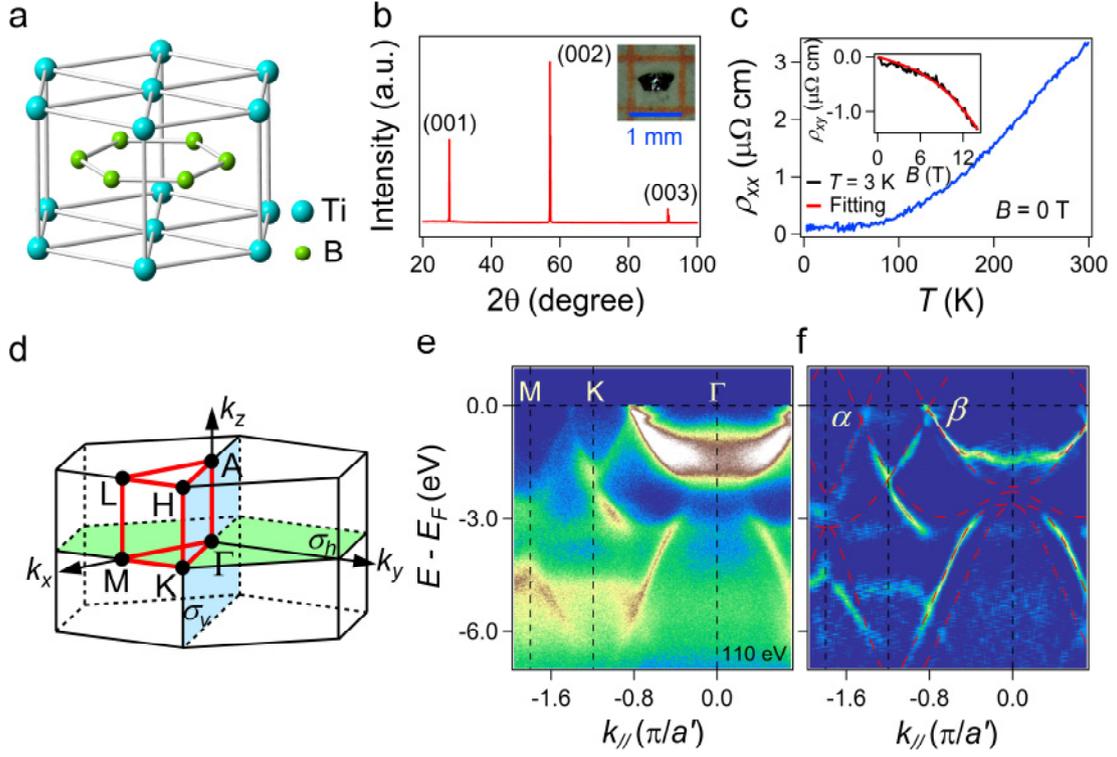

**Fig. 1. Single crystal and wide electronic structure of TiB$_2$.** (a) Crystal structure of TiB$_2$ with space group *P6/mmm* (No. 191). (b) XRD pattern of a TiB$_2$ single crystal. Inset: Picture of a TiB$_2$ crystal against the 1-mm scale. (c) Temperature dependence of the resistivity $\rho_{xx}$ at *B* = 0 T. Inset: Hall resistivity $\rho_{xy}$ (black curve) as a function of magnetic field (*B* ∥ [001]) at *T* = 3 K, with the superimposed fitting result (red curve) using the two-carrier model. (d) 3D bulk BZ with marked high-symmetry points and two mirror planes, $\sigma_h$ (Γ-*K*-*M* plane) and $\sigma_v$ (Γ-*A*-*H* plane). (e, f) ARPES intensity plot and corresponding second derivative plot along the Γ-*K*-*M* direction. *a'* = $\sqrt{3}a/2$ (*a* = 3.0335 Å). The red curves in (f) represent the calculated bulk bands with SOC. The band-crossing features along the *M-K* and *K*-Γ directions are denoted as *α* and *β*, respectively.

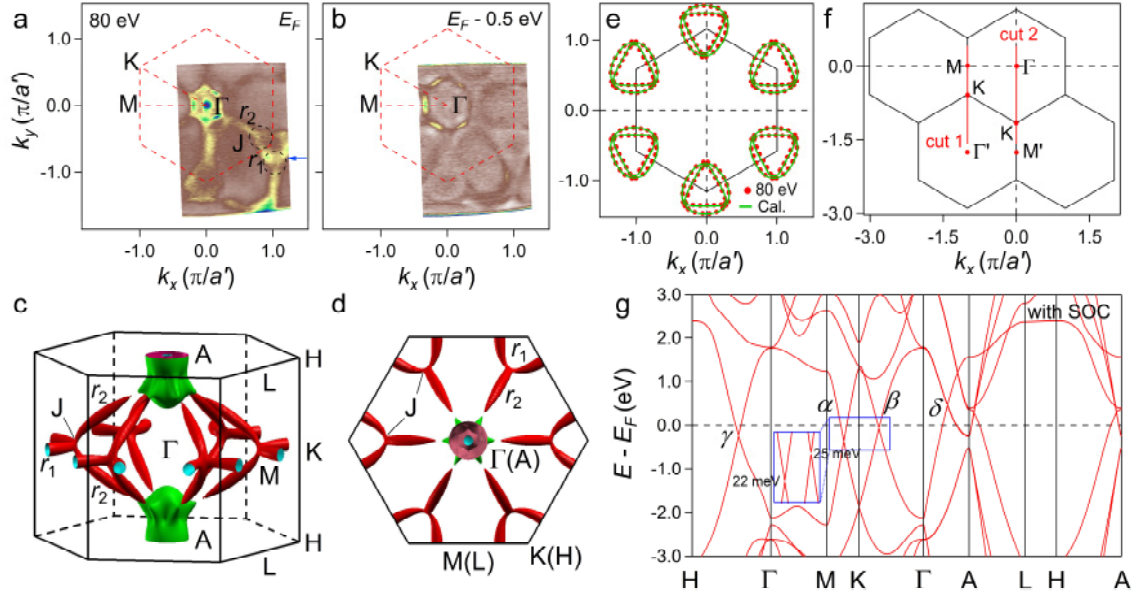

**Fig. 2. FSs and calculated band structure of TiB$_2$.** (a, b) Intensity plots at $E_F$ and 0.5 eV below $E_F$ taken with 80 eV photons. The red dashed lines indicate high-symmetry directions and the first BZ projected on the (001) surface. Two groups of Dirac nodal rings, $r_1$ and $r_2$ (projected on the (001) surface) are resolved. The link points (denoted as $J$ points) of $r_1$ and $r_2$ are indicated by black dashed circles. (c, d) Calculated bulk FSs in the 3D BZ and the top view of the FSs, respectively. The $r_1$ nodal ring surrounding $K$ point is embedded in the Γ-$K$-$M$ plane. The $r_2$ nodal ring surrounding Γ point is embedded in the Γ-$A$-$H$ plane. (e) Comparison of the ARPES data and calculations for $r_1$. The green curves are calculated $r_1$ FSs. The red solid circles represent the experimental data, which are determined as follows, extracting the Fermi wave vectors from 1/3 (marked by a blue arrow) of $r_1$ nodal ring in (a), then rotating it to a whole ring around $K$ point, and last symmetrizing to obtain other rings in the BZ. (f) Schematic 2D BZs with marked cuts 1 and 2 indicating the momentum locations of the measured bands in Fig. 3. (g) Calculated bulk band structure along high-symmetry lines including SOC. Four near-$E_F$ band-crossing features are denoted as $α$, $β$, $γ$, and $δ$, respectively. Inset: The enlarged view of $α$ and $β$, small gaps of 22 and 25 meV open at the nodes respectively, when SOC is taken into consideration.

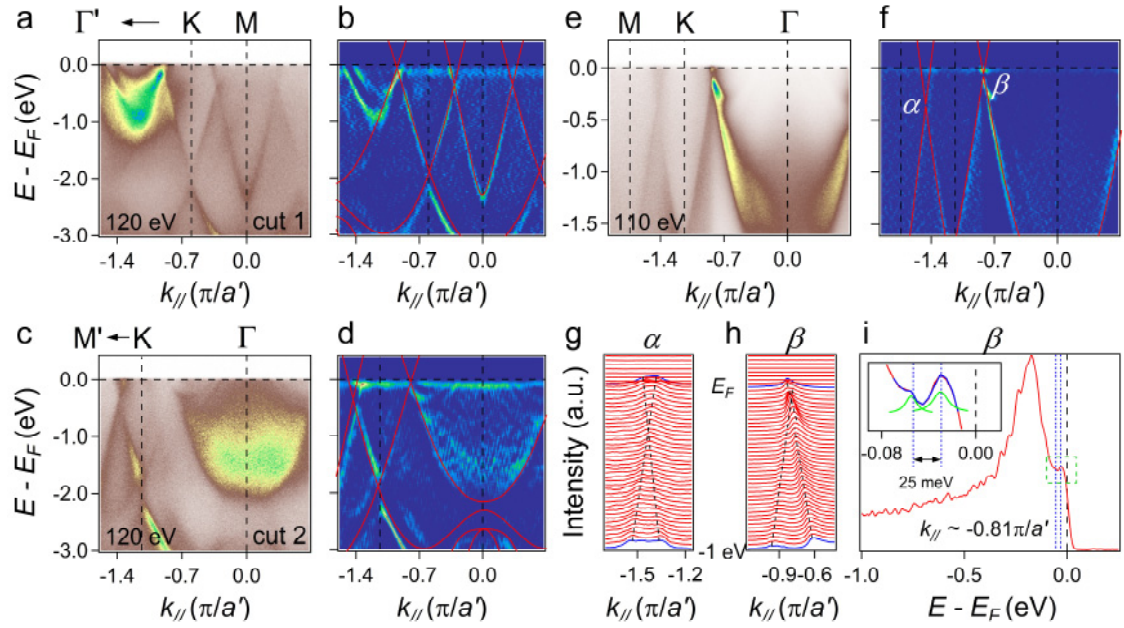

**Fig. 3. Electronic structure of the $r_1$ nodal ring.** (a, b) Intensity plot and corresponding second derivative plot along the *M-K* direction (cut 1 in Fig. 2f) recorded at $h\nu = 120$ eV ($k_z \sim 0.4\pi/c$). (c, d) Same as (a, b), but along the Γ-*K* direction (cut 2 in Fig. 2f). (e, f) Same as (a, b), but recorded along the Γ-*K-M* direction with 110 eV photons ($k_z \sim 0$). The band-crossing features along the *M-K* and *K*-Γ directions, corresponding to the $r_1$ nodal ring surrounding *K* point, are denoted as $\alpha$ and $\beta$ in (f), respectively. The calculated bands with SOC are plotted on the second derivate plots in (b), (d), and (f). (g, h) MDCs plots of $\alpha$ and $\beta$, respectively. The black dashed lines indicate the linear dispersions. (i) EDC taken at the center (~ -0.81 $\pi/a'$) of $\beta$. Inset: Zoomed in EDC near $E_F$ indicated by green dashed rectangle. By fitting the EDC with two Lorentzian profiles, shown as green curves, an opening gap of 25 meV can be determined. The blue curve is the superimposed fitting result.

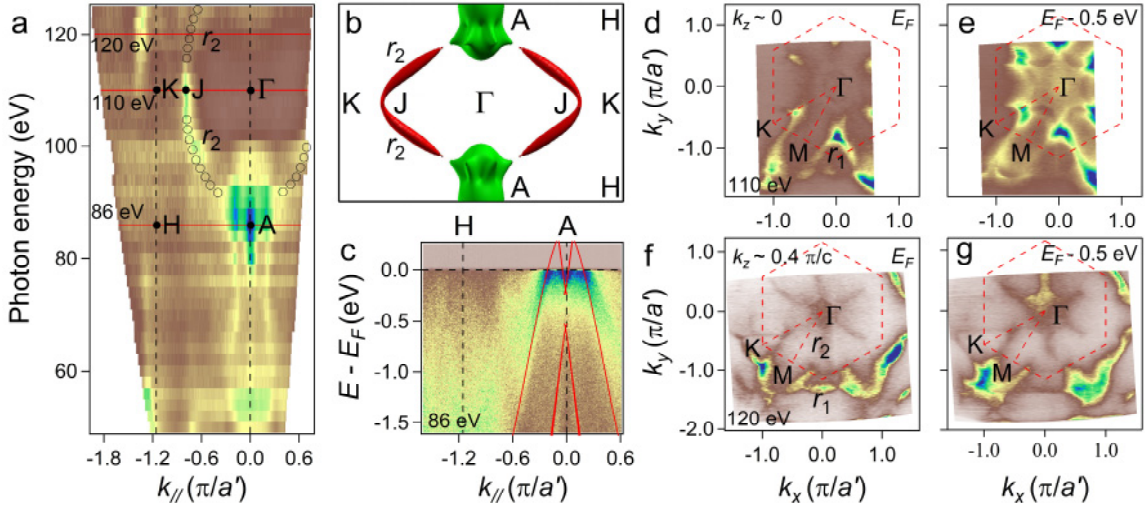

**Fig. 4. Photon-energy-dependent dispersions of TiB$_2$.** (a) Intensity plot in the $h\nu$-$k_{//}$ plane at $E_F$ with $k_{//}$ oriented along the Γ-$K$ ($A$-$H$) direction. The high-symmetry points and nodal-link point are plotted. The $r_2$ nodal ring embedded in the Γ-$A$-$H$ plane surrounding Γ point is indicated by open circles. (b) Projections of calculated bulk FSs on the Γ-$A$-$H$ plane. (c) ARPES spectra taken along the $H$-$A$ line with 86 eV photons ($k_z \sim \pi$). The appended red curves are bulk band calculations with SOC. (d, e) Constant-energy ARPES images at $E_F$ and 0.5 eV below $E_F$ obtained by 110 eV photons ($k_z \sim 0$), respectively. (f, g) Same as (d, e), but taken with 120 eV photons ($k_z \sim 0.4\ \pi/c$).

# Appendix: Materials and Methods

1. **Sample synthesis**

   Single crystals of TiB$_2$ were grown by the Co flux method. The starting elements of Ti (99.995%), B (99.99%) and Co (99.99%) were put into an alumina crucible, with a molar ratio of Ti: B: Co = 1: 2: 4. The mixture was heated up to 1873 K in a high-purity argon atmosphere and then slowly cooled down to 1623 K at a rate of 4 K/h. The TiB$_2$ single crystals were separated from the Co flux using the hot hydrochloric acid solution.

2. **ARPES measurements**

   ARPES measurements were performed at UE112_PGM-2 ARPES end-station of BESSY using photon energies from 50 to 124 eV. The energy and angular resolutions were set to 15 meV and 0.2°, respectively. Samples were cleaved *in situ*, yielding flat mirror-like (001) surfaces. During the measurements, the temperature was kept at 40 K and the pressure was maintained better than 5×10$^{-11}$ Torr.

3. **Band structure calculations**

   First-principles electronic structure calculations on TiB$_2$ were carried out with the projector augmented wave method [38, 39] as implemented in the Vienna *ab initio* Simulation Package [40]. The generalized gradient approximation of Perdew-Burke-Ernzerhof formula [41] was adopted for the exchange-correlation functional. The kinetic energy cutoff of the plane-wave basis was set to be 420 eV. A 20×20×20 *k*-point mesh was utilized for the BZ sampling. The FSs were investigated by adopting the maximally localized Wannier function method [42]. SOC effect is taken into account in all the above calculations.